\documentstyle[aps,epsf,multicol]{revtex}   
\begin{document}
\title{ Weakly bound atomic trimers in ultracold traps}
\author{
M. T. Yamashita$^{(a)}$, T. Frederico$^{(b)}$, Lauro Tomio$^{(c)}$,
and A. Delfino$^{(d)}$
}
\address{$^{(a)}$ Laborat\'orio do Acelerador Linear, Instituto de F\'\i sica
da USP\\ 05315-970, S\~ao Paulo, Brasil \\
$^{(b)}$ Dep. de F\'\i sica, Instituto Tecnol\'ogico de Aeron\'autica,
Centro T\'ecnico Aeroespacial,\\ 12228-900 S\~ao Jos\'e dos Campos, Brasil \\
$^{(c)}$ Instituto de F\'\i sica Te\'orica, Universidade Estadual Paulista,
01405-900 S\~{a}o Paulo, Brasil \\
$^{(d)}$ Instituto de F\'\i sica, Universidade Federal Fluminense,
24210-900 Niter\'oi, RJ, Brasil}
\date{\today}
\maketitle
\begin{abstract}
The experimental three-atom recombination coefficients of the
atomic states $^{23}$Na$|F=1,m_F=-1\rangle$,
$^{87}$Rb$|F=1,m_F=-1\rangle$ and
$^{85}$Rb$|F=2,m_F=-2\rangle$,
together with the corresponding two-body scattering lengths,
allow predictions of the trimer bound state energies for such systems in a trap.
The recombination parameter is given as a function of the weakly bound trimer
energies, which are in the interval $ 1<m(a/\hbar)^2 E_3< 6.9$ for large
positive scattering lengths, $a$. The contribution of a deep-bound state to
our prediction, in the case of $^{85}$Rb$|F=2,m_F=-2\rangle$, for a particular
trap, is shown to be relatively small.
\end{abstract}
\begin{multicols}{2}
The formation of molecules in ultracold atomic traps offers new and
exciting possibilities to study the dynamics of  condensates~\cite{donley}.
It was reported the formation of Rubidium molecules $^{87}$Rb$_2$ in
a bosonic condensate, which allowed  to measure its binding energy with
unprecedented accuracy~\cite{wynar}. Ultracold Sodium molecules
$^{23}$Na$_2$ have also been formed through
photoassociation~\cite{mckenzie}. However,  nothing has been reported till
now about formation of molecular trimers in cold traps.
The first information one is led to ask is the magnitude of
the binding energy of trimers in a cold trap. Two-body scattering
lengths of trapped atoms are well known in several cases, as well as
their closely related dimer binding energy. In the limit of large
scattering lengths, it is necessary to know in addition, one low-energy
three-body observable to predict any other one. In this case, the
detailed form of the two-body interaction is not
important~\cite{ren,nielsen}.
The  recombination rate of three atoms in
the ultracold limit, measured by  atomic losses  in trapped condensed
systems, can supply the necessary information to estimate the trimer
binding energy. For short range interactions, the magnitude of the
recombination rate of three atoms is mainly determined by the two-body
scattering length, $a$~\cite{fedi}.
However, it is important to remark that, still remains a dependence
on one typical low-energy three-body scale~\cite{ren,nielsen}.
Indeed, it is gratifying to note that all the works on three-body
recombination, consistently, present a dependence on a three-body
parameter in addition to the scattering length~\cite{nierec,esryrec,bedarec}.

The aim of the present work is to report on how one can obtain the
trimer binding energy of a trapped atomic system, from the three-body
recombination rate and the corresponding two-body scattering length.
For this purpose, we use a scale independent approach valid in the limit of
large positive scattering lengths (or when the interaction range goes to
zero), obtained from a renormalized zero-range three-body theory~\cite{ren},
which relates the  recombination rate, the scattering length and the
trimer binding energy. Considering the experimental values of the
recombination rates and scattering lengths given in
Refs.~\cite{stamper,burt,soding,roberts}, the method is applied to
predict the trimer binding energies of $^{23}$Na$|F=1,m_F=-1\rangle$,
$^{87}$Rb$|F=1,m_F=-1\rangle$, and $^{85}$Rb$|F=2,m_F=-2\rangle$, where
$|F,m_F\rangle$ is the respective hyperfine states of the total spin $F$.
We note that the bound-states considered here are in fact high-lying
resonances, not true bound states, as they can decay into lower-lying
channels.

The validity of our approach is restricted to sufficiently diluted gases,
because all the scaling relations are derived for three isolated particles.
Also, when the scattering length is tunned via external field in a trap,
the parameters are different from the vacuum values, and consequently
our predictions only apply to that particular experimental conditions.
For the trapped gases that
we are analyzing, the diluteness parameter $\rho a^3$ (where $\rho$ is the
gas density) should not be much larger than one, otherwise one needs to
consider higher order correlations between the particles.
Indeed, we observe that, in general, for the analyzed condensed systems,
the diluteness parameter is much smaller than one. Even in the case of
$^{85}$Rb, where the considered scattering length is obtained
via Feshbach resonance techniques~\cite{roberts}, the diluteness
parameter is about 1/2.

Another relevant remark, pointed out in Ref.~\cite{bra1}, is that the
recombination into deep bound states can affect the theoretical results
that are based on calculation of this rate into shallow states
alone. This additive contribution depends on one more constant, beyond the
three-body scale. However, the fitted contribution of the recombination
into deep bound states, is fortunately much smaller than the contribution of
the shallow bound  state, as found in the case of $^{85}$Rb~\cite{bra1}.
Such evidence supports our estimatives of trimer energies, when $a>0$, that
are obtained by only considering the contribution of recombination rates into
the shallow state.

The values of $a$ are usually defined as large in respect to the
effective range $r_0$, such that $a/r_0>> 1$. The low-energy three-boson
system presents, in this limit, the Efimov effect~\cite{efimov}, where an
infinite number of weakly bound three-body states appears. The size of
such states are much larger than the effective range. The limit $a / r_0
\rightarrow \infty $ can be realized either by $a \rightarrow \infty$
with $r_0$ kept constant or by $r_0 \rightarrow 0$ with $a$ constant.
In the last case, the limit of a zero-range interaction, corresponds to the
Thomas bound-state collapse~\cite{thomas}. In this respect, the Efimov
and Thomas limits are equivalent; or, different aspects of the same
physics~\cite{correl}.
The Thomas-Efimov connection is also reviewed in Ref.~\cite{nielsen}.
In the limit $a/r_0 \rightarrow \infty $,
the details of the interaction for the low-energy three-body system are
contained in one typical three-body scale and the two-body scattering
length (or the dimer bound-state energy, $E_2$); they are enough to
determine all three-body observables~\cite{ren}. Considering, for
example, the trimer binding energy ($E_3$) as the three-body scale, any
three-body observable $({\cal O}_3)$ that has dimension of
[energy]$^\beta$, in the limit of $r_0\rightarrow 0$, can be expressed as
\begin{eqnarray}
{\cal O}_3=E_2^\beta {\cal F}_2(E_2/E_3)=E_3^\beta {\cal F}_3(E_2/E_3) \ .
\label{o3}
\end{eqnarray}
The dimensional factor in front of the above equation (\ref{o3}) is chosen
for convenience as $E_2$ or $E_3$. The scaling function in each case is
${\cal F}_2$ or ${\cal F}_3$.
The existence of the scaling limit for zero range interactions was
verified in Refs.~\cite{amorim,amorim99}. In practice, such limit is
approached by the excited state of the atomic trimer obtained in
realistic calculations, allowing as well the theoretical interpretation
of those excited states as Efimov states~\cite{amorim99}.
Here, we observe that the binding energy $E_3$ refers to the
magnitude of the total energy of the bound-system; the binding energy
with respect to the two-body threshold is defined as $S_3\equiv E_3-E_2$.

The rate of three free bosons to recombine, forming a dimer and one
remaining particle, is given in the limit of zero energy, by the
recombination coefficient\cite{nielsen,fedi}
\begin{eqnarray}
K_3=\frac{\hbar}{m}a^4\alpha ,
\label{k3}
\end{eqnarray}
where $\alpha$ is a dimensionless parameter and $m$ the mass of the atom.
When $a>0$, the recombination
parameter $\alpha$ oscillates between zero and a maximum value, which is
a function of $a$, as shown in Refs.~\cite{nierec} ( $\alpha\le 68.4$ ),
\cite{esryrec} ($\alpha\le 65$) and \cite{bedarec} ($\alpha\le 67.9$).
With amplitude $\alpha_{\rm max}$ and phase $\delta$,
we can write it as \cite{nielsen}
\begin{eqnarray}
\alpha= \alpha_{\rm max}\sin^2\left(1.01\;\ln (a) + \delta\right) \ ,
\label{al}
\end{eqnarray}
where $\delta$ depends on the interaction at short distances. The physics
at short distances, in the three-boson system, is parametrized by one
typical three-body scale, which we have chosen as the unknown trimer
binding energy.  So, by using the general scaling given by
Eq.~(\ref{o3}), one can explicitly express the functional dependence of
$\alpha$ as $\alpha\equiv\alpha\left( \sqrt{{E_2}/{E_3}} \right)$,
considering that, for large scattering lengths we have
$1/a=\sqrt{mE_2/\hbar^2}$.
To exemplify the scaling form of $\alpha$, we can rewrite Eq.(\ref{al}) such
that the
$\sqrt{E_2/E_3}$ dependence is explicit. Therefore,
\begin{eqnarray}
\alpha = \alpha_{\rm max}\sin^2\left(-1.01\ln\sqrt{\frac{E_2}{E_3}} +
\Delta\left(\sqrt{\frac{E_2}{E_3}}\right)\right) \ ,
\label{als}
\end{eqnarray}
where
$\Delta\left(\sqrt{E_2/E_3}\right)=\delta -
1.01\ln\left(\sqrt{{mE_3/\hbar^2}}\right)$.
Our next task is the calculation of the scaling function, by using the
renormalized subtracted Faddeev equations~\cite{ren}.

The three-boson recombination coefficient at zero-energy is derived from
the Fermi's golden-rule as
\begin{eqnarray}
K_3=\frac{2\pi}{\hbar}(2\pi\hbar)^9\int \frac{d^3 p}{(2\pi\hbar)^3}
\left|{\cal T}_{{\rm i}\rightarrow{\rm f}}\right|^2 \delta
\left(\frac{3}{4m}p^2-E_2\right) \ ,
\label{k3int}
\end{eqnarray}
where ${\cal T}_{{\rm i}\rightarrow{\rm f}}$ is the transition amplitude
between the initial (i) and final (f) momentum states, which are normalized as
plane-waves:
$\langle \vec r|\vec p\rangle =$ exp$(-{\rm i}(\vec p/\hbar).\vec r$)
$/(2\pi\hbar)^{3/2}$. The number of atoms $N$ in the condensed state
decreases, due to the recombination process, as:
\begin{eqnarray}
\frac{1}{N}\frac{dN}{dt}=
-\frac{3}{3!} K_3 \rho^2 \ .
\label{l3}
\end{eqnarray}
For each recombination process three atoms are lost, justifying the
factor 3 in the numerator. The factor 3$!$ in the denominator appears
only in case of condensed systems; it counts for
the number of triples in such state~\cite{burt}.

Considering the symmetrized scattering wave-function for the
initial state of three free particles, $| \Phi_0 \rangle =
({1}/{\sqrt{3}})\sum_{i=1}^3|\vec q_i, \vec p_i\rangle$,
we obtain the transition amplitude, in terms of the Faddeev
components of the three body T-matrix, $T_i(E)$, as
\begin{eqnarray}
{\cal T}_{{\rm i}\rightarrow{\rm f}} =
\langle \vec k_i \Phi_b^{(jk)}| [T_j(E)+T_k(E)]
|\Phi_0 \rangle \;\;
,\label{tfa}
\end{eqnarray}
where $(i,j,k)=$ (1,2,3) and cyclic permutations, and $E$ is
the energy of the scattering state is
$E={3q_i^2}/(4m)+p_i^2/m={3}k_i^2/(4m)-E_2$.
$\vec q_i$ is the Jacobi relative momentum of the particle $i$ in respect
to the center of mass of particles $j$ and $k$, $\vec p_i$ is the relative
momentum of $j$ in respect to $k$, and $\vec k_i$ is the relative
momentum of the free particle $i$ in the final state.
$|\Phi_b^{(jk)}\rangle $ is the normalized two-body bound state wave
function of the pair $(jk)$.
The calculation of the Faddeev components is performed with the use of the
subtracted approach given in Ref.~\cite{ren}, such that
\begin{eqnarray}
T_i(E)&=& t_i\left(E-\frac{3q_{i}^2}{4m}\right)\left\{1+
\left(G_0^+(E)-G_0(E_\mu)\right)\times\right.
\nonumber\\
&&\left. \times \left(T_j(E)+T_k(E)\right)\right\} \ ,
\label{tsub}
\end{eqnarray}
where $E_\mu\equiv -\mu^2/m$ is  the subtraction energy scale with
$\mu$ a constant in momentum units. It is possible to vary $\mu$
without changing the physics of the theory as long as the
inhomogeneous term of Eq. (\ref{tsub}) is modified according to
the renormalization group equations~\cite{plb}. $t_i$ is the
two-body $t-$ matrix for the subsystem of particles $(jk)$. For
$E=0$ and zero-range potential the corresponding matrix elements
are given by~\cite{ren}:
\begin{eqnarray}
\langle {\vec p}^\prime |t\left(\frac{-3q^2}{4m}\right)|\vec p
\rangle  =
\tau\left(\frac{-3q^2}{4m}\right)=\frac{1}{m\sqrt{3}\pi^2}\frac{1}
{k-q+{\rm i}\epsilon} \ , \label{t2}
\end{eqnarray}
where $k\equiv\sqrt{4mE_2/3}$.
From  Eqs. (\ref{tsub}) and (\ref{t2}), and for $E=0$,
the matrix elements of
$T_i$ are given by:
\begin{eqnarray}
\langle \vec q_i \vec p_i|T_i(0)|\vec 0 \vec 0\rangle= \tau (0)
\delta(\vec q_i)+2 \tau\left(-{3q^2_i}/{(4m)} \right) h(q_i) \ ,
\label{mtsub}
\end{eqnarray}
where the $s-$wave function $h(q)$ is the solution of
\begin{eqnarray}
h(q)&=&-\frac{\mu^2}{\sqrt{3}\pi^2 kq^2(\mu^2+q^2)}
-\frac{4}{\sqrt{3}\pi} \int^\infty_0 \frac{dq^\prime}{q}
\frac{q^{\prime} h(q^\prime)}{k-q^\prime+{\rm i} \epsilon}\times
\nonumber\\ &\times&
\ln\left(\frac{q^2+q^{\prime2}+qq^\prime}{q^2+q^{\prime2}-qq^\prime}.
\frac{\mu^2+q^2+q^{\prime2}-qq^{\prime}}{\mu^2+q^2+q^{\prime2}+qq^\prime}
\right).\label{h}
\end{eqnarray}

The normalized two-body bound-state wave function, in the zero-range
model, to be introduced in Eq.(\ref{tfa}), is given by
\begin{eqnarray}
\langle \vec p |\Phi_b \rangle =
\frac{1}{\pi}\sqrt{\frac{\hbar}{a}}\frac{1}{({\hbar}/{a})^2+p^2} \ .
\label{phib}
\end{eqnarray}
By considering the above equations, we obtain the final form of the
recombination parameter:
\begin{eqnarray}
\alpha &=& \frac{8(2\pi)^8 m^2}{3\sqrt{3}}\left(\frac{\hbar}{a}\right)^5
\left|{\cal T}_{{\rm i}\to{\rm f}}\right|^2 =
\end{eqnarray}
{\small
\begin{eqnarray}
= 6\sqrt{3}\;(8\pi)^2 \left|1+\frac{16\pi\hbar^2}{3a^2}
\int^\infty_0 dq \frac{q\; h(q)}{k-q+{\rm i}\epsilon}
\ln\left(\frac{k^2+q^2+qk}{k^2+q^2-qk}\right)
\right|^2 \ .\nonumber
\label{k3fim}
\end{eqnarray} }
The numerical results for the recombination parameter are obtained from
the solution of Eq.~(\ref{h}), for different values
of $\mu$. When $\mu\to\infty$, the results approach the scaling
limit~\cite{amorim,amorim99}. Therefore, the theoretical results for
$\alpha$ are shown in Fig.~1 as a function of the ratio $\sqrt{E_2/E_3}$.
The calculations were performed in dimensionless units, such that
all the momentum variables were rescaled in units of $\mu$
(in other words, $\mu=1$ in our calculations).
So, the two-atom binding energy is decreased in respect to this scale.
In that sense, the Thomas-Efimov states appear for $E_2/E_\mu$ going
towards zero, which is equivalent of having $E_2$ fixed and $\mu\to\infty$.
The parameter $\alpha$, shown in Figure 1, is obtained as a function of
the most excited trimer state. We have performed numerical calculations
with at most three Efimov states. The full circles show the results
when exists only one bound state.
When $E_2/E_\mu$ allows two Efimov states, the results are represented
by the solid curve, which is plotted against the energy of the excited
state. With full squares we represent the results
when $E_2/E_\mu$  allows three Efimov states.
The scaling limit is well approached in our calculations. The
maximum $\alpha$ occurs at the threshold ($E_3=E_2$) and
when $(E_3/E_2)^{\frac12}=$0.38~\cite{amorim99}.
So, according to this figure one obtains that $ 1<m(a/\hbar)^2 E_3< 6.9$,
a range consistent with Refs.~\cite{amorim,amorim99,bra}.
The scaling limit for $\alpha$ has been obtained in Refs.
\cite{nierec,bedarec}, but without reference to the weakly bound triatomic
molecular state.

\begin{figure}[thbp]
\setlength{\epsfxsize}{0.8\hsize}
\centerline{\epsfbox{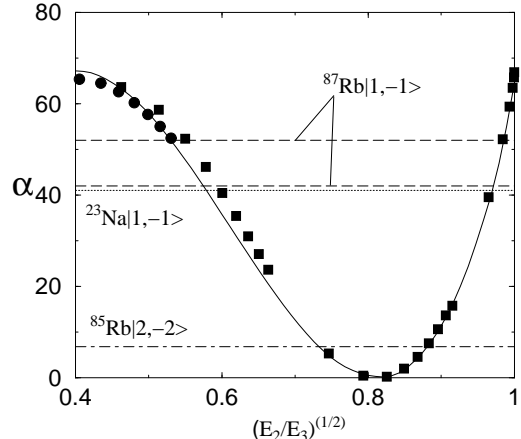}}
\caption[dummy0]{
The dimensionless recombination parameter $\alpha$ as a function of the
ratio between the binding energies of the diatomic and triatomic
molecules. Theoretical results: full circles (one triatomic bound state),
solid line (two triatomic bound states) and full squares
(three triatomic states).
The lines indicate the center of experimental data, given in Table I,
for $^{23}$Na, $^{87}$Rb and $^{85}$Rb. In case of $^{85}$Rb we
subtracted the contribution of the deep bound state, that was
reported in \cite{bra1}.}
\label{fig1}
\end{figure}

In Fig. 1, we represent with horizontal lines the
center of experimental values of $\alpha$, as given in Table I,
for the hyperfine states $^{23}$Na$|1,-1\rangle$,
$^{87}$Rb$|1,-1\rangle$ and $^{85}$Rb$|2,-2\rangle$.
Using the measured values of $\alpha_{expt}$
one obtains, from the universal scaling plot, two weakly bound triatomic
molecular states, denoted by $E_3$ and $E'_3$, consistent with these
values.
Considering the center of experimental values of $\alpha$,
our predicted values for $E_3$ and $E'_3$ are given in Table I
in mK. We are also giving in the table the corresponding known values
of the scattering lengths. The range for the predicted values can be easily
estimated from Fig. 1, considering the corresponding error bars in
$\alpha_{expt}$.
See also Ref.~\cite{weber}, for a recent experiment with ultracold
thermal gas of $^{133}$Cs$|3,3\rangle$,
where the obtained values of $\alpha_{max}$, considering their systematic error limits,
are in good agreement with theory.

One observe that the trap diluteness parameter is smaller than one
in all the cases.
For $^{85}$Rb$|2,-2\rangle$, we study a case corresponding to
$K_3 \approx 3.5\pm 1.5\times 10^{-23}$ cm$^{6}$/s,
extracted from figure 2c of Ref.~\cite{roberts}, measured in an
ultracold non-condensed gas with external field $B=$156 G.
This value  of $B$ corresponds to $a=4000a_0$ ($a_0$ is the Bohr radius)
(See Claussen {\it et al.}~\cite{donley}).
As the resulting value of $\alpha$ is quite small for
$^{85}$Rb$|2,-2\rangle$, one should expect a more
significant contribution from the deep bound state.
Thus, we found instructive to subtract such contribution
from $\alpha_{exp}$, which is about one unit, as found in
Ref.~\cite{bra1}. However, the resulting effect in
the determination of the trimer energy is not so dramatic, as
seen in Fig. 1.
The experimental value of $\alpha$ for $^{87}$Rb$|2,2\rangle$ does not
appear in the figure, as it is well above the maximum.
By increasing the value of $a$ from 5.8 nm to 6.8 nm we can make
the experimental value consistent with our scaling limit approach.
We also point out that the trimer can only support
$E_3$ or $E'_3$, not both simultaneously~\cite{amorim99}.

\end{multicols}

\begin{table}
\caption[dummy0]
{
For the atomic species $^AZ|F,m_F\rangle$, given in the 1st column, we present
in the 6th and 7th columns our predicted trimer binding energies, in respect to the
threshold, $S_3\equiv (E_3-E_2)$ and $S_3^{\prime}\equiv(E'_3-E_2)$, considering
the central values of the experimental dimensionless recombination parameters
$\alpha_{expt}$ (given in the 4th column).
It is also shown the corresponding two-body scattering lengths $a$ (2nd column),
the diluteness parameters $\rho a^3$ (3rd column), and
the dimer binding energies $E_2$ (5th column). For $^{87}$Rb$|1,-1\rangle$, the
recombination process was obtained in Ref.~\cite{burt} for noncondensed ($^*$)
and condensed ($^\dagger$) trapped atoms.
}
\vspace{-0.2cm}
\begin{center}
\begin{tabular}{ccccccc}
$^AZ|F,m_F\rangle$ & $a$(nm) & $\rho a^3$
& $\alpha_{expt}$
&$E_2$ (mK) &
$S_3$(mK) & $S_3^{\prime}$(mK)\\
\hline
$^{23}$Na$|1,-1\rangle$ & 2.75
& 6$\times 10^{-5}$
& 42$\pm$12~\cite{stamper}
& 2.85
& 4.9 & 0.21\\
$^{87}$Rb$|1,-1\rangle$ & 5.8
& 1$\times 10^{-5}$
& 52$\pm$22$^*$~\cite{burt}
& 0.17
& 0.39
& 0.005\\
$^{87}$Rb$|1,-1\rangle$ & 5.8
& 1$\times 10^{-4}$
& 41$\pm$17$^\dagger$~\cite{burt}
&0.17
& 0.30 & 0.013\\
$^{87}$Rb$|2,2\rangle$ &5.8  &
4$\times 10^{-5}$
& 130$\pm$36~\cite{soding}
&0.17
& -    & - \\
$^{85}$Rb$|2,-2\rangle$ &211.6&
0.5
& 7.84$\pm$3.4~\cite{donley,roberts}
&1.3$\times 10^{-4}$
&1.14$\times 10^{-4}$ & 3.8$\times 10^{-5}$
\end{tabular}
\end{center}
\end{table}

\begin{multicols}{2}


In our predictions for the trimer's energies, except for $^{85}$Rb,
we have disregarded the possible much smaller contribution of
the  recombination rate into deep bound states for $a>0$,
considering only recombinations into shallow states.
When the recombination into deep bound states is taken into
account, the curve in Fig. 1 is moved upward by an unknown
amount. But, using the value found in Ref.~\cite{bra1}, this
contribution hardly is going to affect the extracted values for the
trimer's binding energies, given in Table I.
It seems natural that,
if one were to measure the recombination rate as a function of an
applied magnetic field, leading to a Feshbach resonance,
one perhaps could be able to fix this additional contribution and
determine the trimer binding energy~\cite{ref}.
This additional contribution may help to explain part of the
measured value of $\alpha$ for $^{87}$Rb$|2,2\rangle$.

In summary, in the present work, we derived the scaling
dependence of the recombination parameter  as a function of
the ratio  between the  energies of the atomic dimer and the
most excited trimer states. The scaling function tends to a
universal function in the limit of zero-range interaction or
infinite scattering length. The maximum of the recombination rate
comes at the threshold for the appearance of a bound triatomic molecule.
In the cases of diluted gases of $^{23}$Na$|1,-1\rangle$,
$^{87}$Rb$|1,-1\rangle$ and $^{85}$Rb$|2,-2\rangle$,
we use the scaling function, with the corresponding known
experimental values of the recombination rates and two-atom scattering
lengths, to predict for the first time the binding energies of
weakly bound  trimers in ultracold traps.
We stress that the possible contribution of a deep bound state
in our predictions is expected to be relatively small, as
verified for $^{85}$Rb$|2,-2\rangle$, in a particular trap.
We also note that for the $^{85}$Rb$|2,-2\rangle$
the diluteness parameter is about 0.5, a value that may be considered
near the limit of validity of the present approach, which does not
include higher order correlations between the particles.

Finally, we would like to remark that, at a first sight, one
could think that formation of trimers requires four-body collisions,
which are very unlikely unless the density is high.
However, the recent experimental results given in Ref.~\cite{donley}
indicate formation of molecules in the trap, as also discussed in
Ref.~\cite{kohler}.
Therefore, other collision processes like dimer-dimer or
dimer and two atoms could also lead to trimer formations, enhancing the
possibility of producing trimers in a trapped ultracold gas.

We thank V.S. Filho and A. Gammal for discussions; and also the referees for
relevant suggestions. This work is partially supported by Funda\c c\~ao de Amparo
\`a Pesquisa do Estado de S\~ao Paulo and
Conselho Nacional de Desenvolvimento Cient\'{\i}fico e Tecnol\'ogico.

\end{multicols}
\end{document}